\documentclass[twocolumn,pre,twoside,groupedaddress,floatfix,showpacs]{revtex4}

\usepackage{amsmath}
\usepackage{amssymb}
\usepackage{graphicx}
\usepackage{dcolumn}
\usepackage{bm}
\usepackage{color}

\def\d{\text{d}}

\begin{document}


\title{Implications of interface conventions for morphometric thermodynamics}

\author{Andreas Reindl}
\email{reindl@is.mpg.de}
\author{Markus Bier}
\email{bier@is.mpg.de}
\author{S. Dietrich}
\affiliation
{
   Max-Planck-Institut f\"ur Intelligente Systeme, 
   Heisenbergstr.\ 3,
   70569 Stuttgart,
   Germany
}
\affiliation
{
   IV. Institut f\"ur Theoretische Physik,
   Universit\"at Stuttgart,
   Pfaffenwaldring 57,
   70569 Stuttgart,
   Germany
}

\date{15 January, 2015}

\begin{abstract}
Several model fluids in
contact with planar, spherical, and cylindrical walls are
investigated for small number densities within density functional theory. 
The dependence of the solid-fluid interfacial tension on the curvature of
spherical and cylindrical walls is examined and compared with the corresponding 
expression derived within
the framework of morphometric thermodynamics. Particular attention is paid to the
implications of the choice of the interface location, which underlies the definition of the 
interfacial tension. We find that morphometric thermodynamics is never exact for the considered 
systems and that its quality as an approximation depends sensitively on
the choice of the interface location.
\end{abstract}


\maketitle


\section{\label{section:introduction}Introduction}
In recent years morphometric thermodynamics has been used to understand
the influence of geometric constraints on the thermodynamic properties of fluids
\cite{Koenig2004,Laird2012,Evans2014,Goos2007,Roth2006}.
This approach has been motivated by a theorem of integral geometry which is
often referred to as ``Hadwiger's theorem'' \cite{Hadwiger1957,Mecke1998}. 
For example, in Ref.~\cite{Koenig2004} it has been applied to a fluid of hard spheres
bounded by a hard wall. 
Based on the assumption of being an additive, motion-invariant, and continuous
functional of the shape of the walls, the grand canonical potential $\Omega_\text{eq}$
of a confined fluid is given,
in accordance with Hadwiger's theorem, by a linear combination of
only four geometrical measures which characterize the shape $S$ of the confining walls:
volume $\mathcal{V}$, surface area
$\mathcal{A}$, integrated mean curvature $C$, and Euler characteristic $Y$: 
\begin{align}
  \Omega_\text{eq}[S]=-p\mathcal{V}[S]+\gamma_0\mathcal{A}[S]+\kappa
  C[S]+\bar{\kappa}Y[S].
  \label{eq:grand_potential_hadwiger}
\end{align}
According to Eq.~(\ref{eq:grand_potential_hadwiger}) the pressure $p$, the
interfacial tension for a \textit{planar} wall $\gamma_0$, and the
bending rigidities $\kappa$ and $\bar{\kappa}$, unlike the geometrical measures, do not depend on the
shape of the bounding container. This
structure turns morphometric thermodynamics into a very attractive tool, because once the
thermodynamic coefficients $p,\gamma_0,\kappa$, and $\bar{\kappa}$ have been
determined --- preferably by considering simple geometries $S$ --- the grand canonical potential can be 
readily calculated even for systems bounded by complex geometries the shape $S$ of which enters
Eq.~(\ref{eq:grand_potential_hadwiger}) only via the measures $\mathcal{V},\mathcal{A},C$,
and $Y$. This strategy has been used, e.g., in
Refs.~\cite{Evans2014,Goos2007,Roth2006} in order to calculate solvation free energies of
proteins with complex shapes.
\\
As a thermodynamic quantity, which follows from the grand canonical potential,
the interfacial tension $\gamma$ acquires the simple morphometric form \cite{Koenig2004}
\begin{align}
  \gamma=\gamma_0+\kappa\bar{H}+\bar{\kappa}\bar{K},
  \label{eq:surface_tension_koenig}
\end{align}
where $\bar{H}=C/\mathcal{A}$ and $\bar{K}=Y/\mathcal{A}$ denote the averaged
mean and Gaussian curvatures, respectively, of the confining wall \cite{Koenig2004}. In the
case of geometries with constant curvatures, i.e., spherical and
cylindrical walls with radii of curvature $R$, Eq.~(\ref{eq:surface_tension_koenig}) can be
written as
\begin{align}
  \begin{aligned}
    \gamma=
    \begin{cases}
      \gamma_0+\frac{\gamma_1}{R}+\frac{\gamma_2}{R^2},&\text{spherical wall}\\
      \gamma_0+\frac{\gamma_1}{R},&\text{cylindrical wall},
    \end{cases}
  \end{aligned}
  \label{eq:surface_tension_hadwiger_in_curvature}
\end{align}
where $\gamma_{1,2}$ denote coefficients independent of the curvature $1/R$
and where $\gamma_2=0$ in the case of cylindrical walls.

On one hand morphometric thermodynamics has already been applied to several physical
systems confined to geometries with complex shapes \cite{Evans2014,Goos2007,Roth2006} and it has been found to be
in agreement with certain data of hard spheres \cite{Bryk2003,Koenig2004,Laird2012} and of hard rods \cite{Groh1999}
in contact with hard walls.
On the other hand, recently evidences have been provided that a description as in Eq.~(\ref{eq:surface_tension_koenig}),
where the curvature expansion of the interfacial tension for a spherical wall terminates after the quadratic and that of
a cylindrical wall after the linear order in $1/R$, could be incomplete \cite{Blokhuis2013,Urrutia2014,Goos2014}.

Here we propose an explanation for the observation that for certain studies
morphometric thermodynamics appears to be applicable, whereas it is not for others.
To that end, we analyze several model fluids with small number densities in contact with curved
walls using density functional theory (DFT) within the second virial approximation (see 
Sec.~\ref{section:model}). This
technically simple approach allows us to study various kinds of interactions among the fluid particles
(Eq.~(\ref{eq:pp-sw})-(\ref{eq:pp-lj})) with \textit{high}
numerical precision for low densities (see, 
e.g., Fig.~\ref{fig:delta_-0.5_eta_0.02656_comparison_with_laird}).
For this reason we abstain from considering fluids with high densities as they occur, e.g., at two-phase coexistence.
It turns out that those basic features of morphometric thermodynamics we are aiming at reveal themselves already at low 
densities for which one can control the numerical accuracy sufficiently in order to address reliably the corresponding issues.
Moreover we
present exact results for an ideal gas fluid confined by non-hard
spherical and cylindrical walls. We focus
on the interfacial tension $\gamma$ and compare its curvature dependence with the
one predicted from morphometric thermodynamics
(Eq.~(\ref{eq:surface_tension_hadwiger_in_curvature}), see Sec.~\ref{section:discussion}).
It turns out that the morphometric form of the interfacial tension is indeed not valid exactly
and that its quality as an approximation depends on the choice of the location of the
interface underlying the definition of the interfacial tension
(see Sec.~\ref{section:conclusion_and_summary}).



\section{\label{section:model}Model}

We consider a simple fluid composed of particles which interact via an isotropic 
pair potential $U(r)$ which is characterized by an energy scale $U_0$, a length scale 
$L$, and, for technical convenience, a cut-off length $L_c$ such that $U(r>L_c)=0$.
We focus on three distinct types of pair potentials:
\begin{itemize}
   \item the square-well ($U_0<0$) or square-shoulder ($U_0>0$) potential
         \begin{align}
            \beta U(r\leq L_c)=\beta U_0
            \label{eq:pp-sw}
         \end{align}
         with $L=L_c$,

   \item the Yukawa potential ($U_0>0$)
         \begin{align}
            \beta U(r\leq L_c)=\beta U_0
                       \frac{L}{r}\exp\left(-\frac{r}{L}\right),
            \label{eq:pp-yu}
         \end{align}

   \item the Lennard-Jones potential ($U_0>0$)
         \begin{align}
            \beta U(r\leq L_c)=\beta U_0
                       \left(\left(\frac{L}{r}\right)^{12}-\left(\frac{L}{r}\right)^6\right).
            \label{eq:pp-lj}
         \end{align}
\end{itemize}

In the following non-uniform number density profiles $\varrho(\mathbf{r})$ of the fluid in 
contact with walls are studied. To this end density functional theory \cite{Evans1979} offers
a particularly useful approach.
Since the present investigation is focused on \textit{low} densities, we use
the density functional $\Omega[\varrho]$
within the second-virial approximation \cite{HansenMcDonald1976,Schmidt2002}:
\begin{align}
   \beta\Omega[\varrho]&=\int\!\d^3r\,\varrho(\mathbf{r})
                         \left(\ln\left(\varrho(\mathbf{r})L^3\right)-1-\beta\mu+\beta V^\text{ext}(\mathbf{r})\right)
   \label{eq:density_functional}\\
                       &\!\!\!\!+\frac{1}{2}\int\!\d^3r\!\!\int\!\d^3r'\,\varrho(\mathbf{r})\varrho(\mathbf{r'})
                         \left(1-\exp(-\beta U(|\mathbf{r}-\mathbf{r'}|))\right),
   \notag
\end{align}
where $\beta=1/(k_BT)$ is the inverse temperature, $V^\text{ext}(\mathbf{r})$ is the external potential
due to walls, and $\mu$ is the chemical potential $\tilde{\mu}$ shifted by a constant
(with respect to $\mathbf{r}$ and $\varrho$) given by the thermal
de Broglie wavelength $\Lambda$ and the length scale $L$ of the pair potential among the fluid particles:
\begin{align}
   \beta\mu := \beta\tilde{\mu} - 3\ln\left(\frac{\Lambda}{L}\right).
\end{align}

\begin{figure}[!t]
  \includegraphics[width=0.4\textwidth]{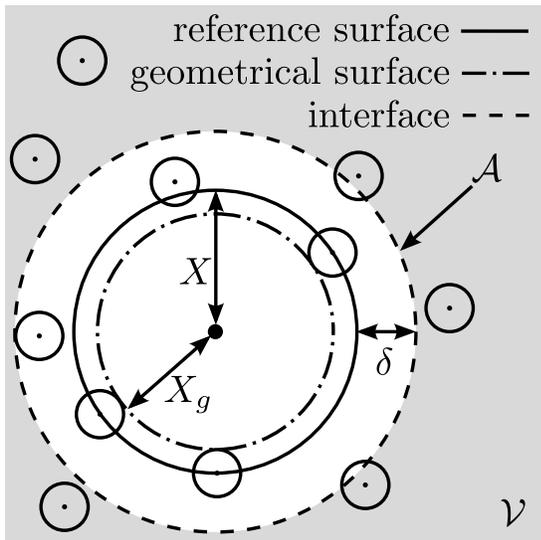}
  \caption{The reference surface (solid line) at position $X$ is determined by
    the external potential. It consists of the set of all reachable locations of the
    \textit{centers} of the fluid particles which are closest to the geometrical wall surface
    (dash-dotted line) at position $X_g$. Depending on the wall-fluid interaction potential,
    the geometrical wall surface and the reference surface can be distinct.
    The interface of area $\mathcal{A}$ (dashed line) --- with respect to which the interfacial tension $\gamma$
    in Eq.~(\ref{eq:surface_tension_with_delta}) is defined --- may differ from
    the reference surface. The parameter $\delta$ measures the offset of the interface
    with respect to the reference surface. This
    sketch refers to the cases of spherical or cylindrical walls for which
    $X\equiv R$. However, the concept involving the parameter $\delta$ is valid also for
    other geometries, in particular for a planar wall with $X\equiv z=0$. If the dashed line runs \textit{within}
    the interior of the reference surface, $\delta$ is taken to be negative. The fluid volume $\mathcal{V}$, which refers to
    the set of all points being not closer to the geometrical wall surface than the interface, is shaded in grey.}
  \label{fig:delta}
\end{figure}

The geometrical properties of the walls enter the description only via the external potential
$V^\text{ext}$. 
In order to determine the thermodynamic coefficients in Eq.~(\ref{eq:grand_potential_hadwiger})
we consider the bulk fluid and the fluid in contact with planar, spherical, and cylindrical 
walls exhibiting constant curvatures.
For these choices the densities in Eq.~(\ref{eq:density_functional}) depend on a single spatial variable only.
\begin{itemize}
\item The uniform \textit{bulk} corresponds to a spatially constant external
      potential, which can be set to zero without loss of generality:
      \begin{align}
         \beta V^\text{ext}(\mathbf{r})=0.
      \end{align}
      As a consequence the equilibrium density $\varrho_\text{eq}(\mathbf{r})=\varrho^\text{bulk}_\text{eq}$ is 
      independent of the position $\mathbf{r}$.

\item A \textit{planar} wall leads to an external potential
      \begin{align}
      \begin{aligned}
         \beta V^\text{ext}(\mathbf{r})=
         \begin{cases}
            \infty,&z<0\\
            \beta V(z),&z\geq0.
         \end{cases}
      \end{aligned}
      \label{eq:planar_external_potential}
      \end{align}
      This potential implies that the equilibrium density $\varrho_\text{eq}(z)$ is identically zero for
      $z<0$. Therefore it is not possible for the \textit{centers} of the fluid
      particles to get closer to the wall than $z=0$. In the following we 
      call the set of accessible points of the \textit{centers} of the fluid particles
      closest to the geometrical wall surface the \emph{reference surface}. 
      In Eq.~(\ref{eq:planar_external_potential}) $V(z)$ represents the excess part of the wall
      potential (i.e., in excess of $V^\text{ext}(z<0)=\infty$)
      as a function of the distance $z$ from the reference surface.
      Depending on the wall-fluid interaction potential, the \textit{g}eometrical wall surface at position $X_g$ in Fig.~\ref{fig:delta}
      and the reference surface at position $X$ in Fig.~\ref{fig:delta}
      can be distinct.

\item A \textit{spherical} wall with radius $R$ of the reference surface is characterized by an
      external potential
      \begin{align}
      \begin{aligned}
         \beta V^\text{ext}(\mathbf{r})=
         \begin{cases}
            \infty,&r<R\\
            \beta V(r),&r\geq R,
         \end{cases}
      \end{aligned}
      \label{eq:spherical_external_potential}
      \end{align}
      where $V(r)$ represents the excess part of the external
      potential as a function of the distance $r$ to the center of the
      sphere.

\item A \textit{cylindrical} wall with radius $R$ of the reference surface is characterized by an
      external potential of the same form as the one in Eq.~(\ref{eq:spherical_external_potential});
      however, in this case, $r$ and $R$ measure distances from the symmetry axis of the cylinder.
\end{itemize}

In accordance with the variational principle underlying density functional theory \cite{Evans1979}
the equilibrium density
$\varrho_\text{eq}(\mathbf{r})$ minimizes the functional in
Eq.~(\ref{eq:density_functional}). The corresponding Euler Lagrange equation
\begin{align}
  \frac{\delta\beta\Omega[\varrho]}{\delta\varrho(\mathbf{r})}\bigg{|}_{\varrho_\text{eq}}=0
\end{align}
is solved numerically. From the equilibrium density profile $\varrho_\text{eq}(\mathbf{r})$,
the grand canonical potential $\Omega_\text{eq}$ follows from \cite{Evans1979} (Eq.~(\ref{eq:density_functional}))
\begin{align}
  \Omega_\text{eq}=\Omega[\varrho=\varrho_\text{eq}].
  \label{eq:grand_canonical_potential}
\end{align}
We have verified that the hard wall sum rule (see, e.g., Ref.~\cite{Bryk2003}) is
fulfilled by the functional in Eq.~(\ref{eq:density_functional}).

\section{\label{section:discussion}Discussion}
The interfacial tension $\gamma$ is defined by 
\begin{align}
  L^2\beta\gamma =
  \frac{\beta\Omega_\text{eq}-\beta\Omega_\text{eq}^\text{bulk}}{\mathcal{A}L^{-2}}=\frac{\beta\Omega_\text{eq}+\beta
  p\;\mathcal{V}}{\mathcal{A}L^{-2}}
  \label{eq:surface_tension}
\end{align}
as the work $\Omega_\text{eq}-\Omega_\text{eq}^\text{bulk}$ per interfacial area $\mathcal{A}$ 
required to create the interface \cite{Rowlinson2002}.
On the right hand side of Eq.~(\ref{eq:surface_tension}) the relation
\begin{align}
  \beta\Omega_\text{eq}^\text{bulk}=-\beta p\;\mathcal{V}
\end{align}
between the grand canonical potential $\Omega_\text{eq}^\text{bulk}$, the fluid volume $\mathcal{V}$, and
the bulk pressure $p$
has been used \cite{Rowlinson2002}.

The wall-fluid interfacial tension $\gamma$ is not an observable
because according to Eq.~(\ref{eq:surface_tension}) its value depends
--- via $\mathcal{V}$ and $\mathcal{A}$ --- on the arbitrary
choice of an interface position.
In order to characterize various interface conventions we introduce a parameter
$\delta$ which measures the offset of a chosen interface position with respect to
the reference surface (see Fig.~\ref{fig:delta}). In addition to the interfacial area $\mathcal{A}$ the
choice of an interface position also determines what is called the fluid volume $\mathcal{V}$, which refers to
the set of all points being not closer to the geometrical wall surface than the interface. As a consequence
$\mathcal{A}$ and $\mathcal{V}$ are functions of $X+\delta$ where $X$ characterizes the reference surface
(see Fig.~\ref{fig:delta}). On the other hand
$\Omega_\text{eq}$ depends on $X$ only, because due to Eq.~(\ref{eq:grand_canonical_potential}) only the parameters of the
substrate potential (i.e., $X$) enter $\Omega_\text{eq}$. Accordingly, one has
\begin{align}
   L^2\beta\gamma(X,\delta) = \frac{\beta\Omega_\text{eq}(X)+\beta
     p\;\mathcal{V}(X+\delta)}{\mathcal{A}(X+\delta)L^{-2}}.
  \label{eq:surface_tension_with_delta}
\end{align}

\begin{figure}[!t]
  \includegraphics[width=0.49\textwidth]{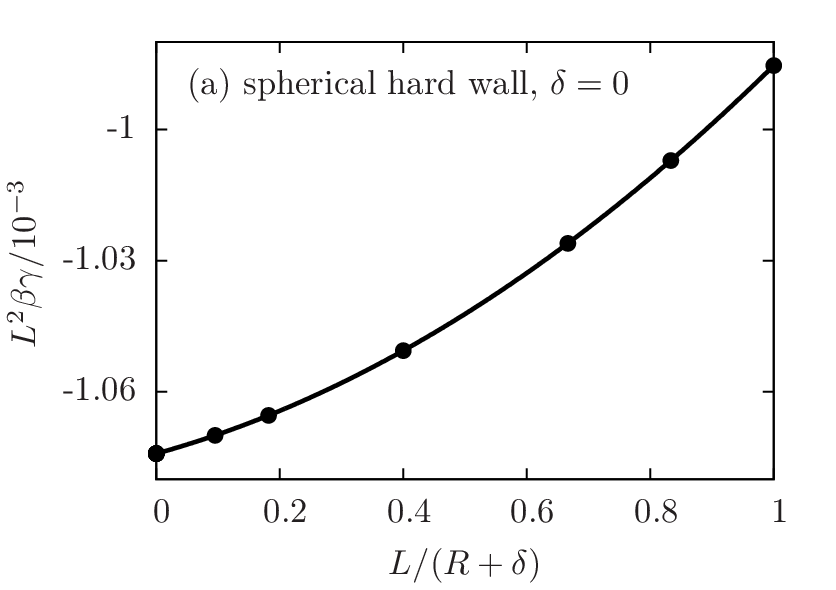}
  \includegraphics[width=0.49\textwidth]{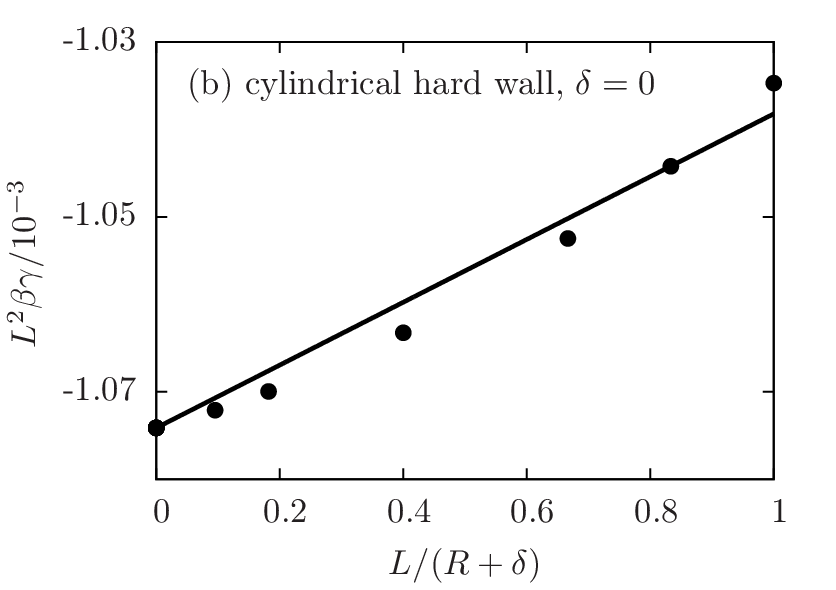}
  \caption{The interfacial tension $\gamma$ of hard spheres with diameter $L$ for the convention $\delta/L=0$ as a function
    of the radius $R$ of the reference surface forming the boundary of a spherical (a) and a cylindrical (b) hard body.
   Polynomials (lines) as predicted by morphometric thermodynamics (up to order $1/(R+\delta)^2$ for spherical walls and up to
   order $1/(R+\delta)$ for cylindrical walls) have been fitted to the numerical data (dots). In panel (a) the fit shows good agreement
   with the data, whereas panel (b) clearly shows deviations from the predicted behavior.
   This means that for the cylindrical wall the curvature expansion of $\gamma$
   Eq.~(\ref{eq:surface_tension_hadwiger_in_curvature}) does not terminate after the first-order term,
   in contradiction to the prediction from morphometric thermodynamics.
   The hard spheres interact among each other via the square-shoulder potential $U(r)$ in
   Eq.~(\ref{eq:pp-sw}) with $\beta U_0=\infty$. The chemical potential is chosen as
   $\beta\mu=-2.768839$ which corresponds to the low packing fraction $\eta=\frac{\pi}{6}\varrho_\text{eq}^\text{bulk}L^3
   \approx0.02656$.}
  \label{fig:delta_0_mu_-2.768839_U0tilde_1e300_Lc_1._pp_sw}
\end{figure}

Our data are calculated within the convention $\delta/L=0$ as this choice is
convenient for various types of interactions $U$.
Moreover, in certain studies (see, e.g., Ref.~\cite{Henderson2002}) this choice has been argued to be 
``the natural one from the point of view of statistical mechanics''.
Figure~\ref{fig:delta_0_mu_-2.768839_U0tilde_1e300_Lc_1._pp_sw} shows the
interfacial tension $\gamma$ for hard spheres in contact with spherical 
(Fig.~\ref{fig:delta_0_mu_-2.768839_U0tilde_1e300_Lc_1._pp_sw}(a)) and
cylindrical (Fig.~\ref{fig:delta_0_mu_-2.768839_U0tilde_1e300_Lc_1._pp_sw}(b))
walls with various radii $R$ of the reference surface.
The packing fraction
$\eta=\pi\varrho_\text{eq}^\text{bulk}L^3/6$ is chosen sufficiently small such that the
second virial approximation is valid. The plot for the cylindrical walls 
(Fig.~\ref{fig:delta_0_mu_-2.768839_U0tilde_1e300_Lc_1._pp_sw}(b)) reveals a non-linear increase, similar to the 
case of spherical walls (Fig.~\ref{fig:delta_0_mu_-2.768839_U0tilde_1e300_Lc_1._pp_sw}(a)). 
This means that the curvature expansion of $\gamma$ in terms of powers of $1/R$ does not
terminate after the first-order term, in contradiction to the prediction from morphometric thermodynamics for cylindrical walls.

Evaluating Eq.~(\ref{eq:surface_tension_with_delta}) for $\delta=0$ and arbitrary $\delta\neq0$ and
exploiting that $\Omega_\text{eq}$ does not depend on $\delta$ leads to
\begin{align}
  \begin{aligned}
    L^2\beta\gamma(X,\delta)=&\frac{\mathcal{A}(X)}{\mathcal{A}(X+\delta)}L^2\beta\gamma(X,\delta=0)\\
                               &+\frac{L^2\beta
                                 p}{\mathcal{A}(X+\delta)}\left(\mathcal{V}(X+\delta)-\mathcal{V}(X)\right).
  \end{aligned}
  \label{eq:transformation_of_surface_tension}
\end{align}
According to Eq.~(\ref{eq:transformation_of_surface_tension}) the interfacial tension
$\gamma(X,\delta=0)$ calculated for the convention $\delta/L=0$ can
be translated to that for any other choice of the convention.
For example the convention $\delta/L=-0.5$ is often used when discussing hard spheres confined by hard walls
because this choice renders the interface to coincide with the geometrical wall surface at position $X_g$
in Fig.~\ref{fig:delta}, which, in this case, is separated from the reference surface at position $X$ in
Fig.~\ref{fig:delta} by a distance given by the particle radius $L/2$.

\begin{figure}[!t]
  \includegraphics[width=0.49\textwidth]{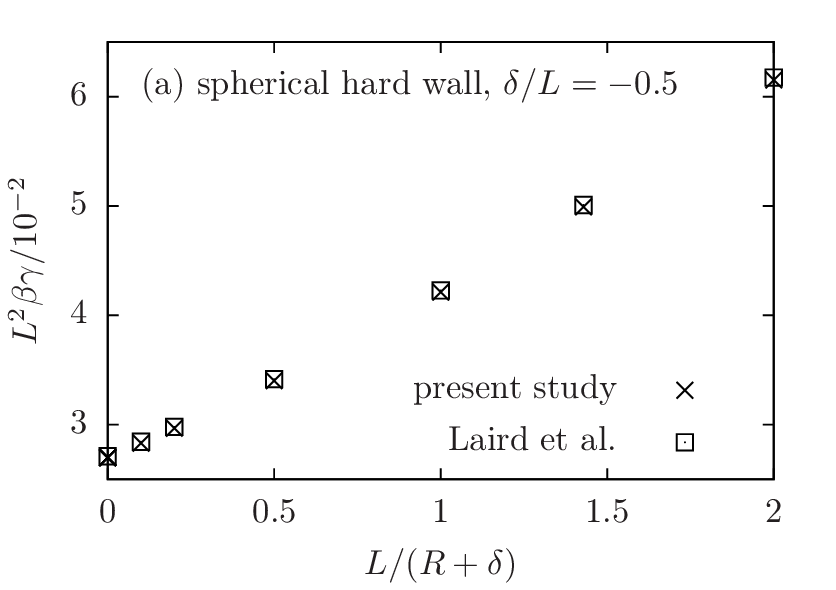}
  \includegraphics[width=0.49\textwidth]{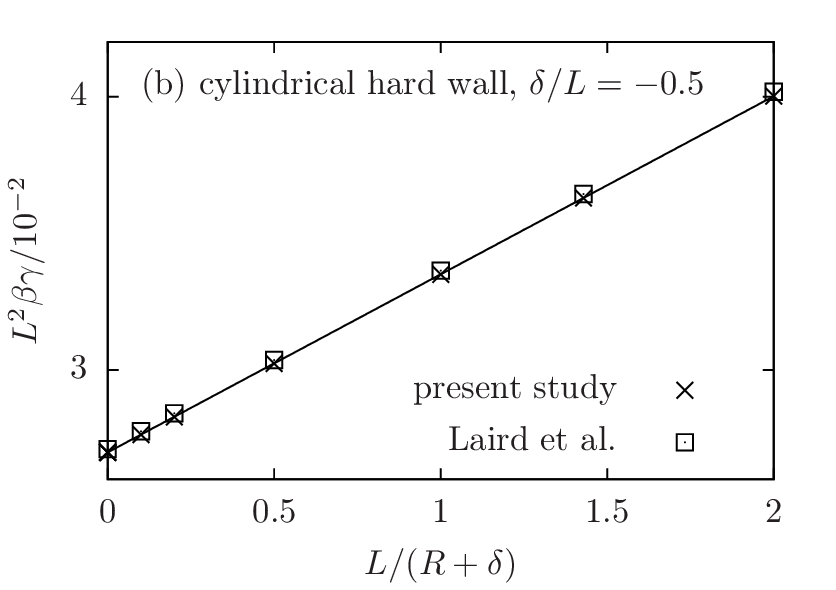}
  \caption{Comparison of the simulation results of Laird et
    al. \cite{Laird2012} (squares $\Box$) with the corresponding data of the present DFT study
    (crosses $\times$) for the packing
    fraction $\eta\approx0.02656$. The latter have been obtained by translating the data displayed in
    Fig.~\ref{fig:delta_0_mu_-2.768839_U0tilde_1e300_Lc_1._pp_sw} (for $\delta=0$) into the convention
    $\delta/L=-0.5$ by using
    Eq.~(\ref{eq:transformation_of_surface_tension}). In contrast to
    Fig.~\ref{fig:delta_0_mu_-2.768839_U0tilde_1e300_Lc_1._pp_sw}(b), the plot of
    the interfacial tension in the case of a cylindrical wall (b) almost coincides
    with a straight line.}
  \label{fig:delta_-0.5_eta_0.02656_comparison_with_laird}
\end{figure}

In Fig.~\ref{fig:delta_-0.5_eta_0.02656_comparison_with_laird} simulation
results of Laird et al. \cite{Laird2012} for hard spheres with packing
fraction $\eta=0.02656$, obtained for $\delta/L=-0.5$, are plotted together
with the data of Fig.~\ref{fig:delta_0_mu_-2.768839_U0tilde_1e300_Lc_1._pp_sw} which
have been translated into the convention $\delta/L=-0.5$ according to Eq.~(\ref{eq:transformation_of_surface_tension}).
The agreement with the simulation data is very good.
In particular, within this convention for $\delta$, in the case of a cylindrical wall 
(Fig.~\ref{fig:delta_-0.5_eta_0.02656_comparison_with_laird}(b)) the data points almost coincide with a straight line,
in accordance with the prediction of morphometric thermodynamics. This finding has also been
confirmed for the packing fractions $\eta\approx0.053$ and
$0.101$, for which the
respective plots are qualitatively similar to those in
Fig.~\ref{fig:delta_-0.5_eta_0.02656_comparison_with_laird} except that, as
expected, the results of the present virial expansion deviate more and more from the
simulation data upon increasing the density.

Figures~\ref{fig:delta_0_mu_-2.768839_U0tilde_1e300_Lc_1._pp_sw} and
\ref{fig:delta_-0.5_eta_0.02656_comparison_with_laird}, which are based on the
same microscopic system, show that the interfacial tension depends strongly
on the choice of the convention for $\delta$.
Upon varying $\delta$ not only the sign of $\gamma$ may change, as already
noted in Ref.~\cite{Bryk2003}, but also the magnitude and even the qualitative functional form, 
which is revealed clearly in the case of cylindrical walls.

\begin{figure}[!t]
  \includegraphics[width=0.49\textwidth]{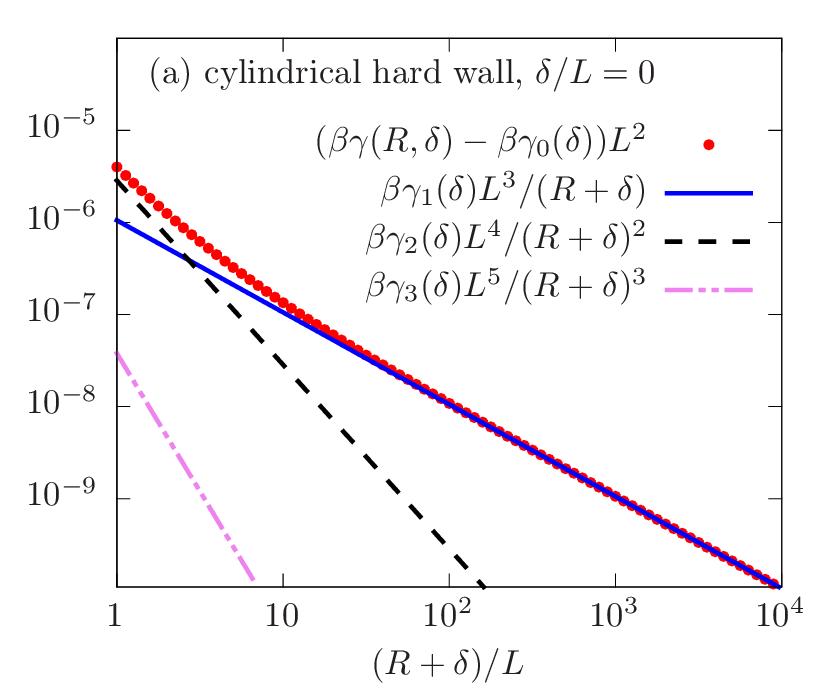}
  \includegraphics[width=0.49\textwidth]{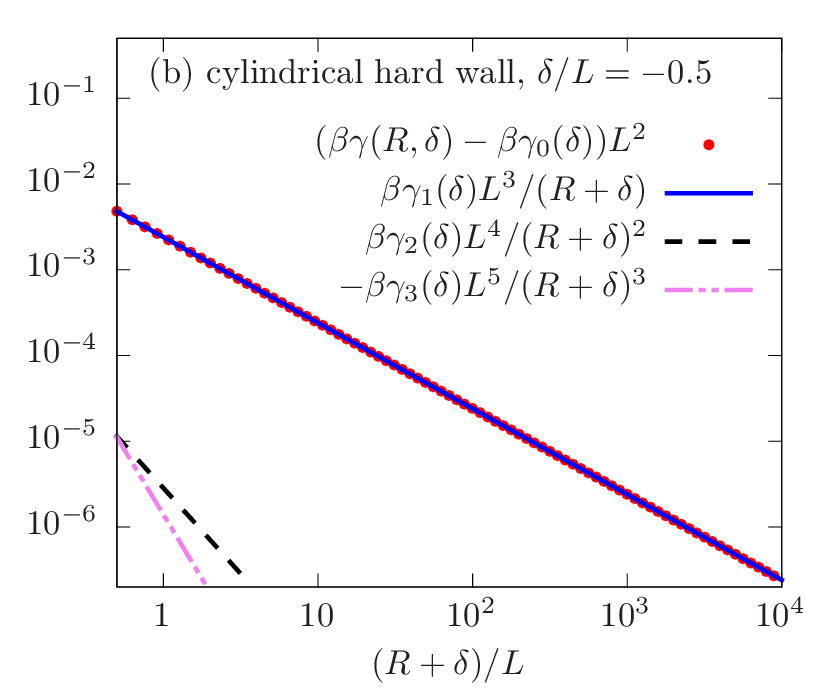}
  \caption{Dependence of the interfacial tension $\gamma$ for hard spheres at a cylindrical
    hard wall on the radius $R$ of the reference surface and on the
    shift parameter $\delta$. The contribution of the planar wall $\gamma_0(\delta)$, which
    corresponds to the limit $R\rightarrow\infty$, has been subtracted.
    The numerical data (red dots) are obtained within the convention $\delta/L=0$
    (a) and translated to the convention $\delta/L=-0.5$ (b) via
    Eq.~(\ref{eq:transformation_of_surface_tension}). The three straight lines represent the
    terms $n=1,2,3$ of the curvature expansion in Eq.~(\ref{eq:curvature_expansion}). In the
    case $\delta/L=0$ the coefficient $\gamma_2(\delta)$ is the largest one, i.e., for small $R$ the contribution
    $\beta\gamma_2(\delta)L^4/(R+\delta)^2$ is the dominant one so that a
    deviation from the leading behavior $\sim 1/(R+\delta)$ becomes obvious for $R/L\lesssim3$. Within the
    convention $\delta/L=-0.5$ the leading coefficient $\gamma_1(\delta)$ is the largest one, i.e., the leading term
    $\beta\gamma_1(\delta)L^3/(R+\delta)$ is the dominant one.
    Here $U(r)$ is given by Eq.~(\ref{eq:pp-sw}) with $U_0\rightarrow\infty$ and
    $\beta\mu=-3.88$ so that $\eta\approx0.01$. Note that at $(R+\delta)/L=1$
    the plots render the values of the dimensionless coefficients $\beta\gamma_n(\delta)L^2$.}
  \label{fig:mu_-3.88_U0tilde_1e300_Lc_1._pp_sw}
\end{figure}

Figure~\ref{fig:mu_-3.88_U0tilde_1e300_Lc_1._pp_sw} shows the behavior of the
same microscopic system as above (hard spheres exposed to a hard cylindrical wall)
described in terms of two
conventions for $\eta\approx0.01$. The data are presented in log-log plots which facilitate the
identification of power laws in $1/(R+\delta)$. In this presentation the contribution
$\gamma_0(\delta)$ of the planar wall is subtracted so that the
plotted quantity vanishes for $R\rightarrow\infty$. Within the
convention $\delta/L=0$, at $R\approx 3L$ there is a crossover between two power laws.
Thus the dependence of the interfacial tension on $1/R$
consists of more than the leading term $\sim 1/R$ which, according to morphometric
thermodynamics, would be the
only one allowed for the cylindrical configuration. 
The behavior is different within the convention $\delta/L=-0.5$.
There, within this presentation, the interfacial tension is represented by an almost straight
line throughout the whole range of $(R+\delta)/L$ shown.
In order to analyze the curvature dependence of the interfacial tension more
quantitatively, we assume that $\gamma(R,\delta)$ can be expanded
in terms of a power series in $1/(R+\delta)$:
\begin{align}
  L^2\beta\gamma(R,\delta)=L^2\sum\limits_{n=0}^\infty L^n\frac{\beta\gamma_n(\delta)}{(R+\delta)^n}.
  \label{eq:curvature_expansion}
\end{align}
For various radii $R$ of the reference surface of the
curved wall, the interfacial tension $\gamma$ is calculated numerically (red dots in
Fig.~\ref{fig:mu_-3.88_U0tilde_1e300_Lc_1._pp_sw}) and fitted to the curvature
expansion Eq.~(\ref{eq:curvature_expansion}) with $n\leq 10$. This way the
coefficients $\gamma_n(\delta)$ have been determined. Only the coefficient
$\gamma_0(\delta)$, which is the interfacial tension at a planar wall, can be
obtained independently without fitting. In
Fig.~\ref{fig:mu_-3.88_U0tilde_1e300_Lc_1._pp_sw} the terms
$L^{n+2}\beta\gamma_n(\delta)/(R+\delta)^n$
corresponding to $n\in\{1,2,3\}$ in the curvature expansion of Eq.~(\ref{eq:curvature_expansion}) are
plotted as lines. Within the
convention $\delta/L=0$ the quadratic coefficient $\gamma_2$, which vanishes within
morphometric thermodynamics, is even larger than the linear
coefficient $\gamma_1$; this explains the crossover between two
power laws describing the numerical data (red dots). However, within the convention $\delta/L=-0.5$
the first-order coefficient $\gamma_1$ is much larger than the higher order coefficients
$\gamma_n, n\geq2$, so that in this case
morphometric thermodynamics is a very good approximation of the exact
curvature dependence of the interfacial tension.
Within the class of systems with square-well-like or square-shoulder-like
particle-particle interactions, we have studied a large variety of configurations
within the convention $\delta/L=0$
in the same way as shown in
Fig.~\ref{fig:mu_-3.88_U0tilde_1e300_Lc_1._pp_sw}.
Fluids with packing fractions $\eta\in\{0.01,\,0.02,\,0.05,\,0.10\}$ 
and with interaction strengths $\beta
U_0\in\{-0.1,0.1,\,1,\,\infty\}$ have been examined near spherical and cylindrical hard
walls. In addition to the interfacial tension $\gamma$, the dimensionless
excess adsorption $\Gamma$ \cite{Rowlinson2002}
\begin{align}
  L^2\Gamma(X,\delta)=\frac{N(X)-\varrho_\text{eq}^\text{bulk}\,\mathcal{V}(X+\delta)}{\mathcal{A}(X+\delta)L^{-2}},
\end{align}
has been calculated for $U_0>0$ where $N(X)=\int_{\mathcal{V}(X)}d^3r\varrho(\mathbf{r})$ denotes the number
of fluid particles. Our corresponding observations can be summarized as follows:
\begin{itemize}
  \item Apart from opposite signs, both the excess adsorption $\Gamma$ and the interfacial tension $\gamma$
    exhibit a similar dependence on the radius of curvature $R$.
  \item The functional form of
    $L^2\left(\beta\gamma(R,0)-\beta\gamma_0(0)\right)$ is similar when comparing
    fluids with the same bulk state near
    a spherical and a cylindrical hard wall. 
    Because in all considered cases the third order coefficient $\gamma_3$ is smaller than 
    $\gamma_1$ and $\gamma_2$, morphometric thermodynamics turns out to be
    a better approximation for spherical than for cylindrical walls.
  \item Decreasing packing fractions $\eta$ or interaction strengths $U_0>0$
    result in a shift of the crossover between the power laws
    describing $L^2(\gamma(R,0)-\gamma_0(0))$
    towards larger radii $R$, i.e., the
    second coefficient $\gamma_{2}$ becomes larger in comparison to the
    first coefficient $\gamma_{1}$. Apart from
    that, the behavior of $L^2\left(\beta\gamma(R,0)-\beta\gamma_0(0)\right)$ is similar to the one shown in
    Fig.~\ref{fig:mu_-3.88_U0tilde_1e300_Lc_1._pp_sw}. For $U_0<0$,
    $L^2\left(\beta\gamma(R,0)-\beta\gamma_0(0)\right)$ exhibits a zero because the
    coefficients $\gamma_{1}$ and $\gamma_{2}$ have opposite signs.
  \item For some of the above systems the data have been translated to the
    convention $\delta/L=-0.5$ by using
    Eq.~(\ref{eq:transformation_of_surface_tension}). In these cases, for the spherical wall configurations
    the third order coefficient $\gamma_{3}$ is much smaller than $\gamma_1$ and $\gamma_2$, and for
    the cylindrical wall configurations the leading coefficient $\gamma_{1}$ is
    much larger than the subleading ones. Therefore, within this convention morphometric thermodynamics
    turns out to be a very good approximation.
\end{itemize}

The comparison of the plots in Figs.~\ref{fig:mu_-3.88_U0tilde_1e300_Lc_1._pp_sw}(a) and (b)
shows that the coefficients $\gamma_n(\delta)$ in the curvature expansion
(Eq.~(\ref{eq:curvature_expansion})) indeed depend on the chosen convention $\delta$.
In order to examine the implications of shifting the position of the
interface (see Fig.~\ref{fig:delta}) this dependence is
investigated more closely.
The derivative of
Eq.~(\ref{eq:surface_tension_with_delta}) with respect to $\delta$ for fixed $R$ leads to
\begin{align}
  \begin{aligned}
    L\frac{\partial}{\partial\delta}L^2\beta\gamma(R,\delta)&=-L^3\beta p-d\frac{L}{R+\delta}L^2\beta\gamma(R,\delta)\\
    d&=
     \begin{cases}
       2,&\text{spherical wall}\\
       1,&\text{cylindrical wall},
     \end{cases}
  \end{aligned}
  \label{eq:derivative_of_betagamma_definition}
\end{align}
where $\mathcal{V}'(R+\delta)=-\mathcal{A}(R+\delta)$ and $\mathcal{A}'(R+\delta)/\mathcal{A}(R+\delta)=d/(R+\delta)$ have been used.
The derivative of
Eq.~(\ref{eq:curvature_expansion}) with respect to $\delta$ gives
\begin{align}
  \begin{aligned}
    L\frac{\partial}{\partial\delta}L^2\beta\gamma(R,\delta)=L^3\sum\limits_{n=0}^\infty
     \bigg\{&L^n\frac{\beta\gamma'_n(\delta)}{(R+\delta)^n}\\
     &-L^nn\frac{\beta\gamma_n(\delta)}{(R+\delta)^{n+1}}\bigg\}.
  \end{aligned}
  \label{eq:derivative_of_betagamma_expansion}
\end{align}
Equating Eqs.~(\ref{eq:derivative_of_betagamma_definition}) and (\ref{eq:derivative_of_betagamma_expansion}) and using
Eq.~(\ref{eq:curvature_expansion}) leads to
\begin{align}
  \begin{aligned}
    -L^3\beta p&=L^3\beta\gamma_0'(\delta)+\sum\limits_{n=1}^\infty
     \frac{L^n}{(R+\delta)^n}\\
      &\times\left\{L^3\beta\gamma'_n(\delta)+(d-n+1)L^2\beta\gamma_{n-1}(\delta)\right\}
  \end{aligned}
  \label{eq:equalling_of_derivatives}
\end{align}
for all $R$. Comparison order by order in $(R+\delta)^{-1}$ in Eq.~(\ref{eq:equalling_of_derivatives}) renders
\begin{align}
  O((R+\delta)^0):\quad-L^3\beta p=L^3\beta\gamma'_0(\delta)
  \label{eq:differential_equation_0}
\end{align}
and
\begin{align}
  \begin{aligned}
    O((R+\delta)^{-n}),n\geq1:&\\
    L^3\beta\gamma'_n(\delta)+&(d-n+1)L^2\beta\gamma_{n-1}(\delta)=0.
  \end{aligned}
  \label{eq:differential_equation_n}
\end{align}
Integration of Eqs.~(\ref{eq:differential_equation_0}) and (\ref{eq:differential_equation_n}) with respect to $\delta$ leads to
the following iterative algorithm for determining the dependence of
the coefficients $\gamma_n(\delta),n\geq0$, on $\delta$:
\begin{align}
  \begin{aligned}
    &n=0:\\
    &L^2\beta\gamma_0(\delta)=L^2\beta\gamma_0(0)-L^3\beta p\frac{\delta}{L},\\
    &n\geq1:\\
    &L^2\beta\gamma_n(\delta)=L^2\beta\gamma_n(0)+(n-d-1)\int\limits_0^\delta\frac{\text{d}\tilde{\delta}}{L}
      L^2\beta\gamma_{n-1}(\tilde{\delta})\\
    &d=
     \begin{cases}
       2,&\text{spherical wall}\\
       1,&\text{cylindrical wall}.
     \end{cases}
  \end{aligned}
  \label{eq:coefficients_iteration}
\end{align}

\begin{figure}[!t]
  \includegraphics[width=0.49\textwidth]{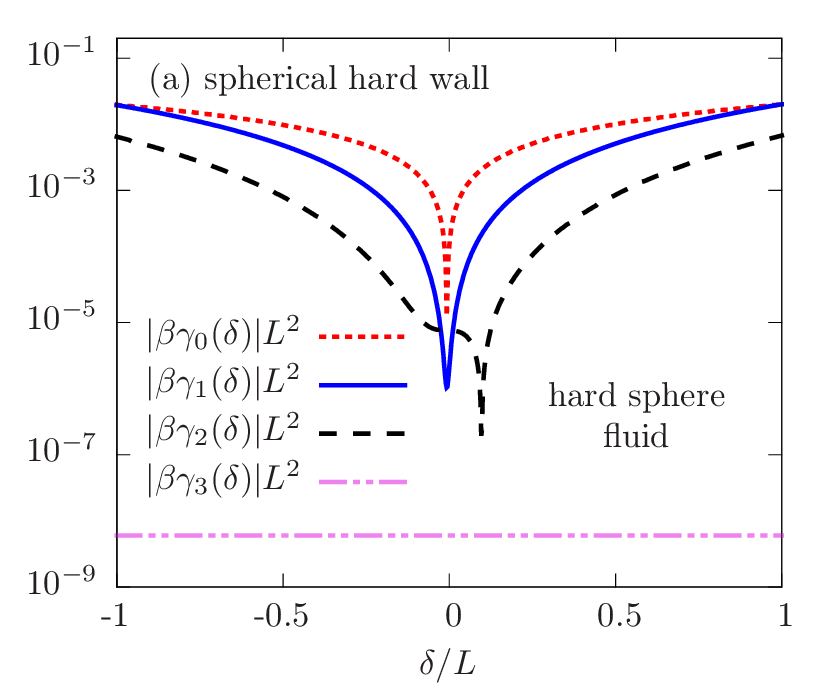}
  \includegraphics[width=0.49\textwidth]{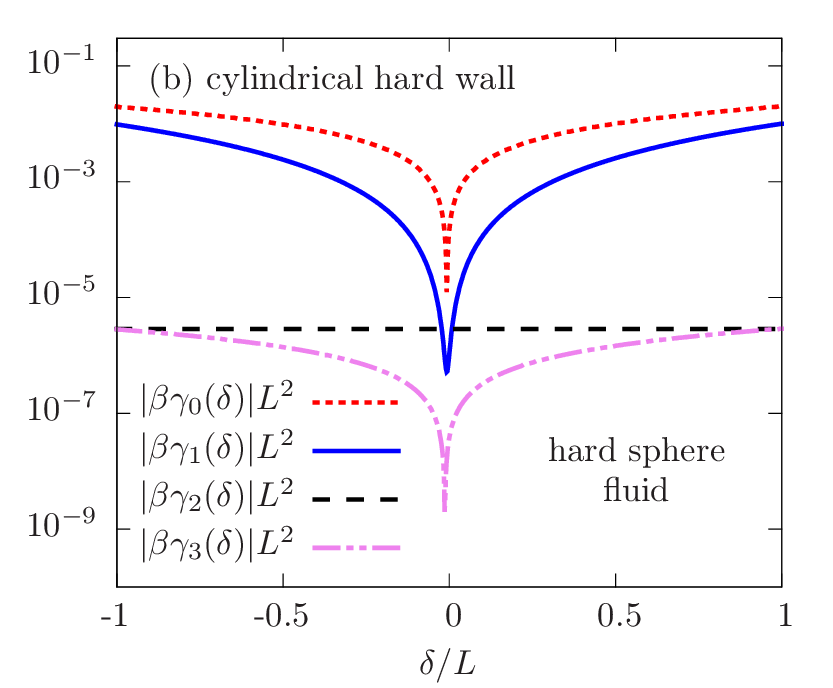}
  \caption{Reduced coefficients $\gamma_n(\delta), n\in\{0,1,2,3\},$ characterizing the curvature expansion
    in Eq.~(\ref{eq:curvature_expansion}) as function of the shift parameter
    $\delta$ and as obtained from Eq.~(\ref{eq:coefficients_iteration}). The data correspond to a hard sphere
    fluid at $\beta\mu=-3.88$, so that $\eta\approx0.01$, exposed to hard spherical (a) or cylindrical (b) walls.}
  \label{fig:intdspl_mu_-3.88_U0tilde_1e300_Lc_1._pp_sw}
\end{figure}

\begin{figure}[!t]
  \includegraphics[width=0.49\textwidth]{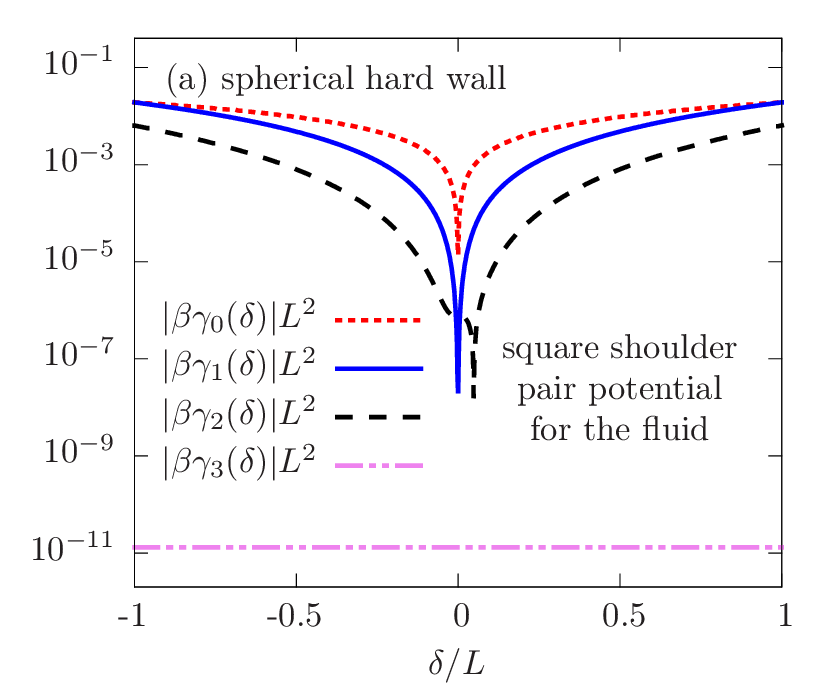}
  \includegraphics[width=0.49\textwidth]{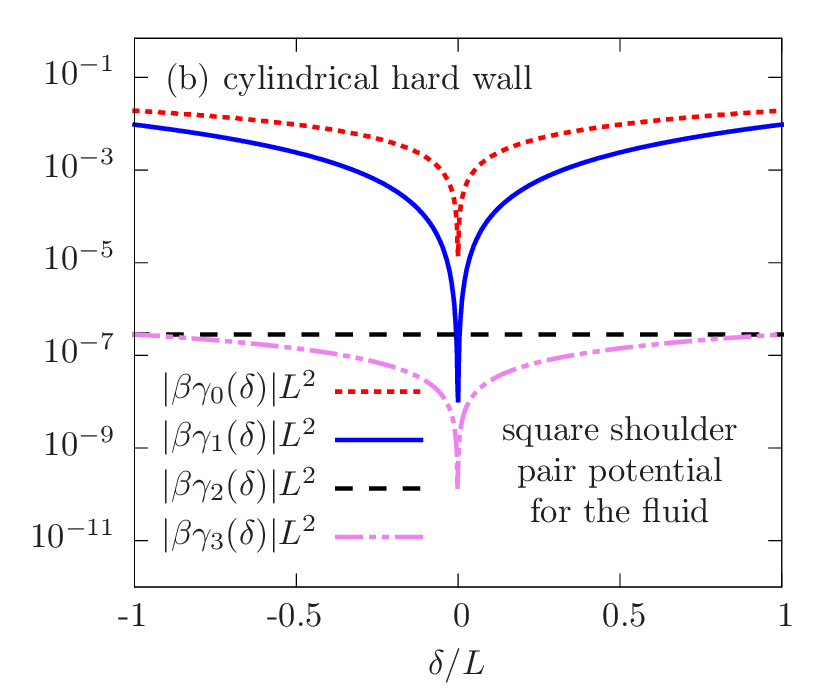}
  \caption{Reduced coefficients $\gamma_n(\delta), n\in\{0,1,2,3\},$ characterizing the curvature expansion
    in Eq.~(\ref{eq:curvature_expansion}) as function of the shift parameter
    $\delta$ and as obtained from Eq.~(\ref{eq:coefficients_iteration}). The data
    correspond to a fluid with a square-shoulder pair potential $U(r)$ (Eq.~(\ref{eq:pp-sw})) with $\beta U_0=0.1$ 
    and $\beta\mu=-3.95$ so that $\eta\approx0.01$. The fluid is exposed to hard spheres (a) or hard cylinders (b).}
  \label{fig:intdspl_mu_-3.95_U0tilde_0.1_Lc_1._pp_sw}
\end{figure}

\begin{figure}[!t]
  \includegraphics[width=0.49\textwidth]{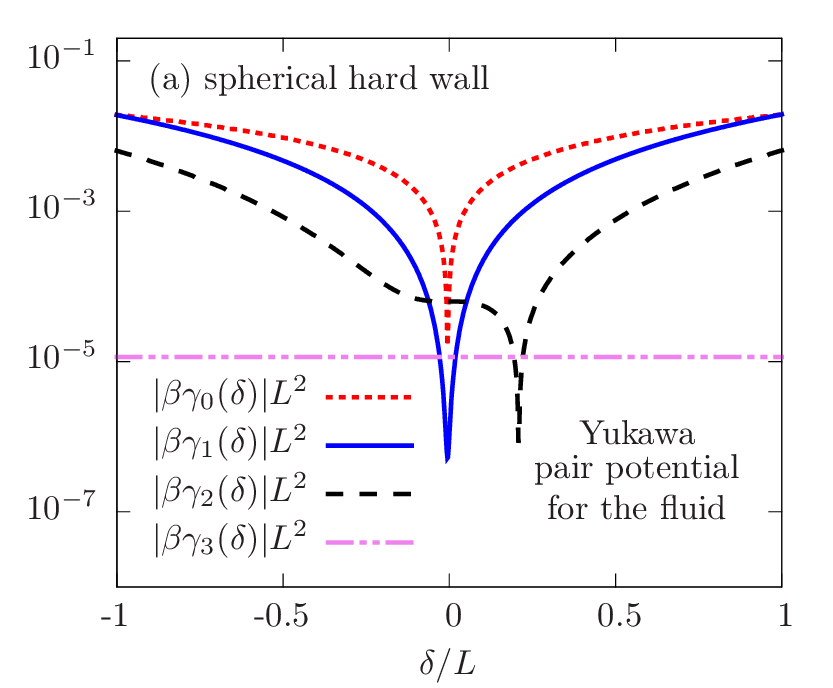}
  \includegraphics[width=0.49\textwidth]{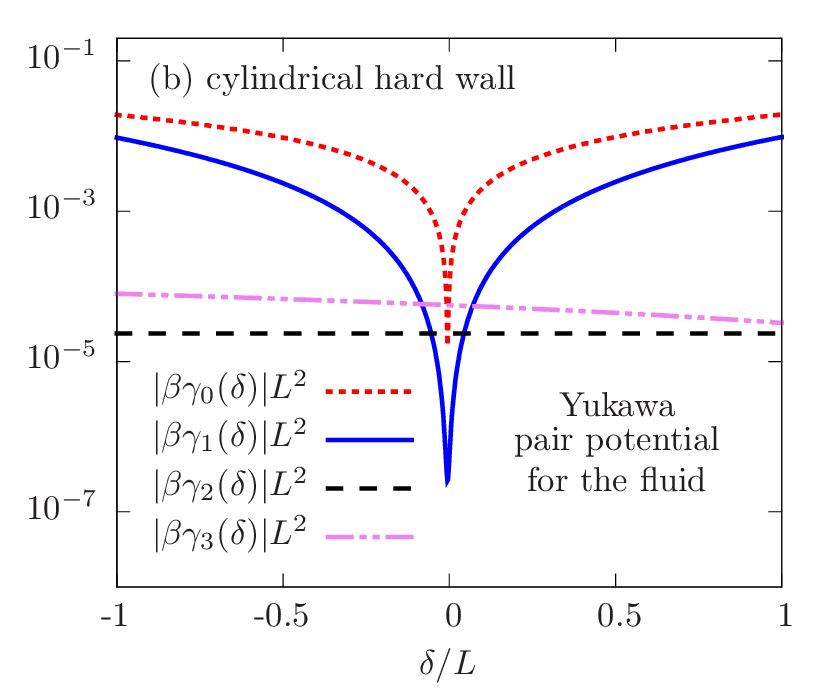}
  \caption{Same as Fig.~\ref{fig:intdspl_mu_-3.95_U0tilde_0.1_Lc_1._pp_sw} for a fluid with a Yukawa
    pair potential (Eq.~(\ref{eq:pp-yu})) with $\beta U_0=0.1$,
    $\beta\mu=-3.94$, and $L_c/L=5$ so that $\eta\approx0.01$.}
  \label{fig:intdspl_mu_-3.94_U0tilde_0.1_Lc_5._pp_yu}
\end{figure}

\begin{figure}[!t]
  \includegraphics[width=0.49\textwidth]{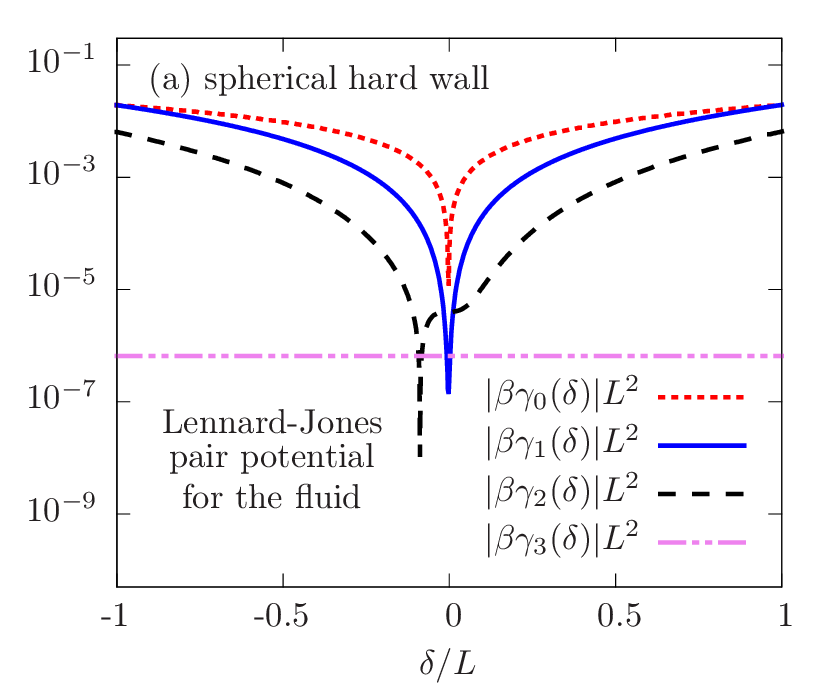}
  \includegraphics[width=0.49\textwidth]{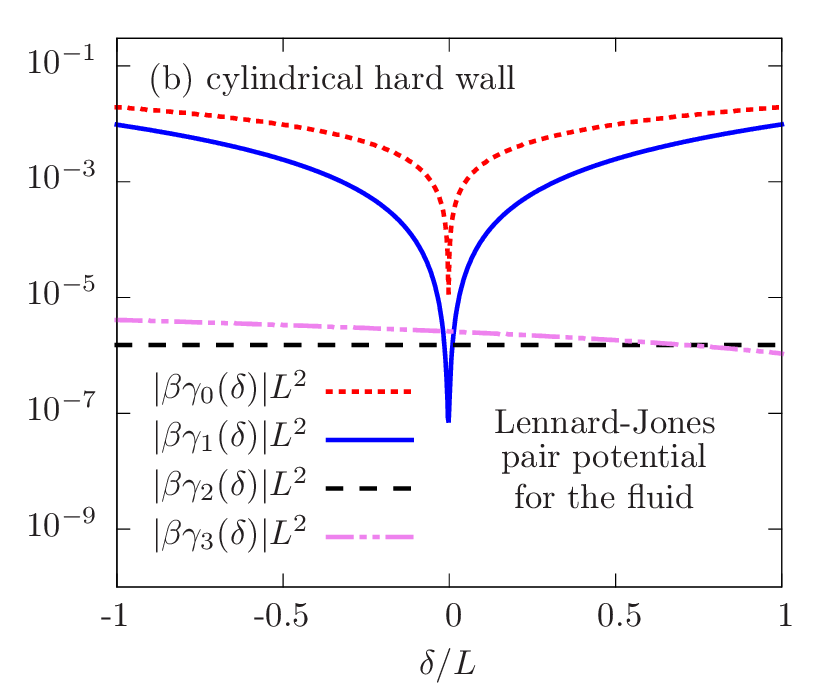}
  \caption{Same as Fig.~\ref{fig:intdspl_mu_-3.94_U0tilde_0.1_Lc_5._pp_yu} for a fluid with a
    Lennard-Jones pair potential (Eq.~(\ref{eq:pp-lj})) with $\beta U_0=0.1$,
    $\beta\mu=-3.92$, and $L_c/L=5$ so that $\eta\approx0.01$.}
  \label{fig:intdspl_mu_-3.92_U0tilde_0.1_Lc_5._pp_lj}
\end{figure}

\begin{figure*}[!t]
  \includegraphics[width=0.49\textwidth]{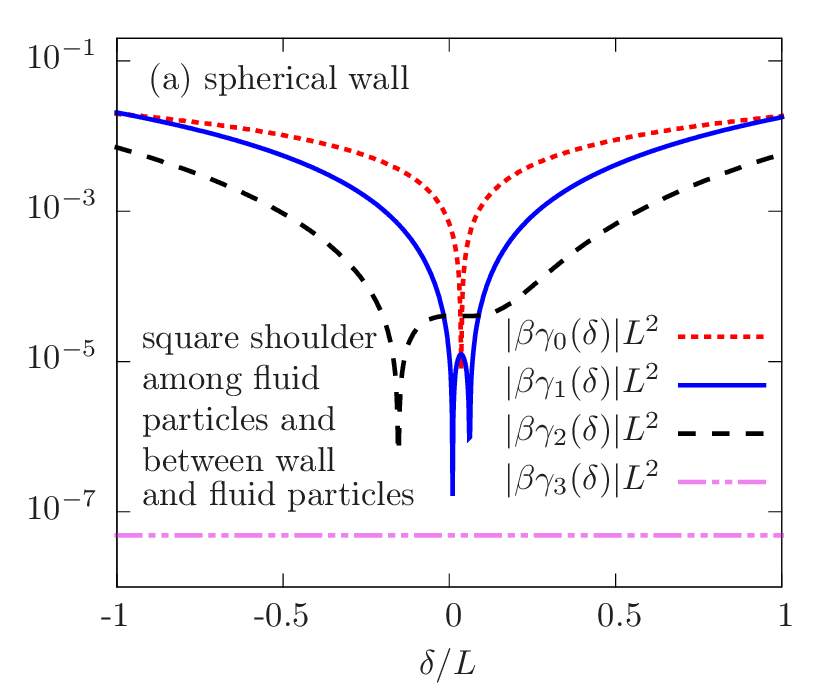}
  \includegraphics[width=0.49\textwidth]{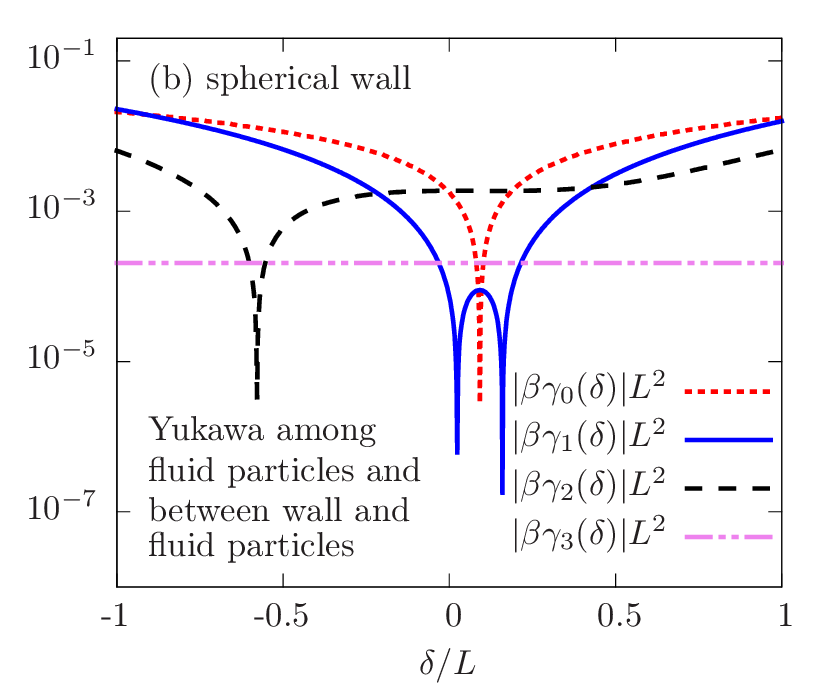}
  \includegraphics[width=0.49\textwidth]{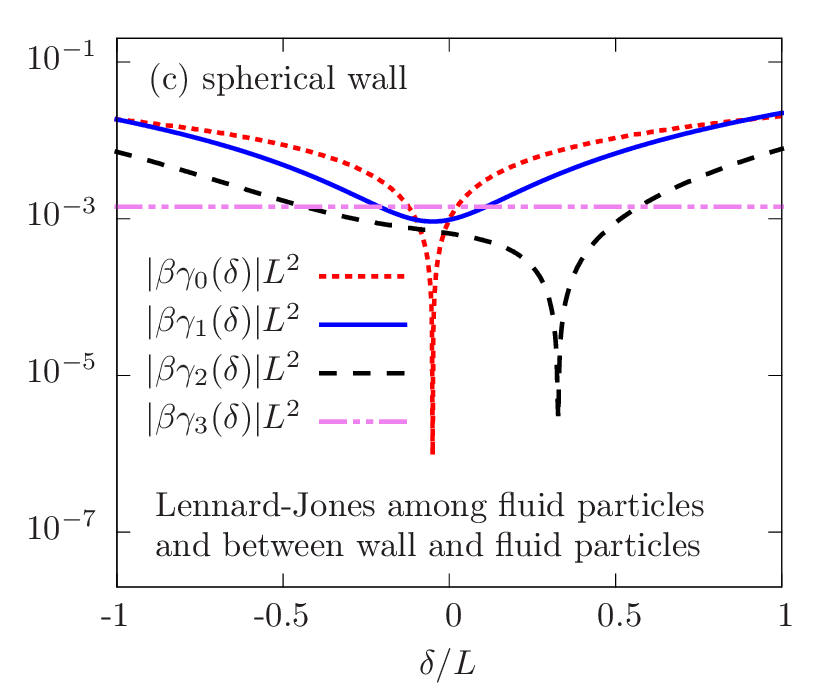}
  \caption{Reduced coefficients $\gamma_n(\delta), n\in\{0,1,2,3\},$ characterizing the curvature expansion
    in Eq.~(\ref{eq:curvature_expansion}) as function of the shift parameter
    $\delta$ and as obtained from Eq.~(\ref{eq:coefficients_iteration}). Panels (a),
  (b), and (c) belong to the same systems as in
  Figs.~\ref{fig:intdspl_mu_-3.95_U0tilde_0.1_Lc_1._pp_sw}(a),
  \ref{fig:intdspl_mu_-3.94_U0tilde_0.1_Lc_5._pp_yu}(a), and
  \ref{fig:intdspl_mu_-3.92_U0tilde_0.1_Lc_5._pp_lj}(a), respectively, with an additional
  soft part $V$ of the external substrate potential: (a) $V$ given by
  Eq.~(\ref{eq:soft_part_sw}), (b) $V$ given by Eq.~(\ref{eq:soft_part_yu}), and (c)
  $V$ given by Eq.~(\ref{eq:soft_part_lj}). The data correspond to the following choice of system
  parameters: $\beta V_p^{sq}=\beta V_p^{yu}=\beta V_p^{LJ}=0.1$ (see Appendix \ref{sec:soft_parts}),
  $L_w/L=1$, and $L_{cw}/L=5$.}
  \label{fig:intdspl_U0tilde_0.1_Vp_0.1}
\end{figure*}

The dependence of the coefficients $\gamma_n(\delta)$ on $\delta$ is fully
determined once in Eq.~(\ref{eq:coefficients_iteration})
the values of $\gamma_{n'}(0)$ for all $n'\leq n$ are known. Here the values $\gamma_n(0)$ are
obtained by fitting Eq.~(\ref{eq:curvature_expansion}) for $n\leq10$ to the numerical data within
the convention $\delta/L=0$. Considering terms in Eq.~(\ref{eq:curvature_expansion}) to such high orders was necessary
in order to achieve a sufficiently high precision for the actually interesting coefficients $\gamma_1(\delta),\dots,\gamma_3(\delta)$
(see Figs.~\ref{fig:mu_-3.88_U0tilde_1e300_Lc_1._pp_sw}-\ref{fig:intdspl_U0tilde_0.1_Vp_0.1}); taking the additional terms of order 
$n\geq4$ into account guarantees that
these coefficients $\gamma_1(\delta),\dots,\gamma_3(\delta)$ are not
affected by the fast-decaying contributions of the full curvature expansion.
Thereby we have found that the 
ratio of the leading coefficients for the spherical wall, $\gamma_1^s(0)$, 
and for the cylindrical wall, $\gamma_1^c(0)$, takes the value $\gamma_1^s(0)/\gamma_1^c(0)=2$ 
for all systems considered here, with a relative deviation of
$10^{-7}$ or less. Comparison of that numerical result with the exact relation $\gamma_1^s(\delta)/\gamma_1^c(\delta)=2$, which
follows from the fact that the total curvatures $J=1/R_1+1/R_2$ of a sphere, $J^s$,
and of a cylinder, $J^c$, are related by $J^s=2J^c$ (see Refs.~\cite{Koenig2004,Blokhuis2013}),
validates our numerical approach.
Motivated by this relationship between
the leading coefficients we have considered also the ratio of the next-to-leading
coefficients. For small packing fractions we have obtained
$\gamma_2^s(\delta)/\gamma_2^c(\delta)\approx8/3$ for $\delta=0$,
independently of the particle-particle interaction potential $U$. The value of this ratio does
depend on the convention for $\delta$ because the spherical coefficient 
$\gamma_2^s(\delta)$ varies with $\delta$, whereas the cylindrical coefficient $\gamma_2^c$ 
is constant (see discussion below). Actually, the value $8/3$ can be obtained from the exact analytical expressions
for the surface tension in the
low-density limit for a fluid of \textit{hard spheres} \cite{Urrutia2014}.
Moreover, the exact expression in Eq.~(\ref{eq:second_order_coefficients_ideal_gas})
(see Appendix \ref{sec:ideal_gas}) describes the deviation of the ratio $\gamma_2^s(0)/\gamma_2^c(0)$
from $8/3$ for an ideal gas of particles as function of the strength $\beta
V_p^{yu}$ of a short-ranged excess external potential in addition to the hard wall potential.

It is interesting to pay special attention to the
coefficient $\gamma_{n=d+1}$ in Eq.~(\ref{eq:coefficients_iteration}) which
is the coefficient of the lowest order being not in accordance with morphometric thermodynamics.
For this order $n=d+1$ one has  $n-d-1=0$ and
therefore $\gamma_{d+1}$ is constant in $\delta$ (see Eq.~(\ref{eq:coefficients_iteration})
for $n\geq1$). This checks with
Fig.~\ref{fig:mu_-3.88_U0tilde_1e300_Lc_1._pp_sw} (corresponding to $d=1$), where
the values of the coefficients $\gamma_2$ can be read from the lines $\sim1/(R+\delta)^2$ at $(R+\delta)/L=1$.
The value of $\gamma_2$ is not vanishing and it is the same in both conventions for $\delta$. 
This implies that within morphometric thermodynamics the curvature expansion is not exact for any 
convention for $\delta$. On the other hand, if, as a consequence of
Eq.~(\ref{eq:coefficients_iteration}), the $R$-dependence of $\gamma(R,\delta)$ within morphometric 
thermodynamics would be exact for any single convention for $\delta$, it would be
exact for all conventions for $\delta$. 
However, this statement is of no practical use, because, on the basis of numerical data,
it is virtually impossible to prove that there is a convention for $\delta$ within which
the morphometric form of the interfacial tension is \textit{exact}.
In contrast, Fig.~\ref{fig:mu_-3.88_U0tilde_1e300_Lc_1._pp_sw} shows that even if the
non-morphometric coefficients $\gamma_n(\delta)$ ($n\geq2$ for a cylindrical wall) are numerically small for
one convention for $\delta$ (see 
Fig.~\ref{fig:mu_-3.88_U0tilde_1e300_Lc_1._pp_sw}(b)) they may be large for
another (see Fig.~\ref{fig:mu_-3.88_U0tilde_1e300_Lc_1._pp_sw}(a)).
The reason for this is that the operation of approximating
the curvature-dependence of the interfacial tension by the form 
predicted within morphometric thermodynamics does not commute with the operation of
shifting the interface.

Figure~\ref{fig:intdspl_mu_-3.88_U0tilde_1e300_Lc_1._pp_sw} shows the dependence
of the coefficients $\gamma_n(\delta)$, $0\leq n\leq3$, of the curvature expansion
Eq.~(\ref{eq:curvature_expansion}) on the shift parameter $\delta$. Except for
the first non-morphometric coefficient (i.e., $\gamma_3$ for a sphere and $\gamma_2$ for a cylinder),
which is constant in $\delta$, the coefficients
$\gamma_n(\delta)$ vary over several orders of magnitude upon changing $\delta$.
This is particularly pronounced near $\delta/L=0$,
where the morphometrically allowed coefficients are small; in the case of 
a cylindrical wall $\gamma_1$ is even smaller than the leading non-morphometric
coefficient $\gamma_2$. However, apart from this region around $\delta/L=0$,
e.g., at $\delta/L=-0.5$, the morphometrically allowed coefficients
exceed the leading non-morphometric coefficient by several orders of magnitude. These
observations are in agreement with the findings of
Fig.~\ref{fig:mu_-3.88_U0tilde_1e300_Lc_1._pp_sw} which is based on the same system and where for each 
convention for $\delta$ the coefficients, the values of which are rendered at $(R+\delta)/L=1$, have been
fitted independently. 
Figure~\ref{fig:intdspl_mu_-3.88_U0tilde_1e300_Lc_1._pp_sw} demonstrates
that the interfacial tension cannot be represented exactly by the form obtained within morphometric 
thermodynamics, and that the quality of the approximation of the interfacial tension by the morphometric form
depends on the position of the interface parameterized by the shift $\delta$.

So far we have mainly focussed on hard sphere fluids near hard walls. In the following
we discuss to which extent the aforementioned observations can be extended to other systems. 
This will be discussed along the lines of Fig.~\ref{fig:intdspl_mu_-3.88_U0tilde_1e300_Lc_1._pp_sw}.

Figure~\ref{fig:intdspl_mu_-3.95_U0tilde_0.1_Lc_1._pp_sw} displays the corresponding
results for a fluid which is governed by a square-shoulder pair
potential of finite strength $\beta U_0$, acting as a representative for interaction potentials of finite range. 
The comparison with Fig.~\ref{fig:intdspl_mu_-3.88_U0tilde_1e300_Lc_1._pp_sw} does not reveal
qualitative changes.
Figures~\ref{fig:intdspl_mu_-3.94_U0tilde_0.1_Lc_5._pp_yu} and
\ref{fig:intdspl_mu_-3.92_U0tilde_0.1_Lc_5._pp_lj}, respectively,
display the corresponding results for a fluid with a Yukawa pair potential,
representing exponentially decaying interaction
potentials, and for a Lennard-Jones pair potential, representing algebraically decaying pair potentials.
Although comparing them with Figs.~\ref{fig:intdspl_mu_-3.88_U0tilde_1e300_Lc_1._pp_sw} and
\ref{fig:intdspl_mu_-3.95_U0tilde_0.1_Lc_1._pp_sw} reveals certain differences, the main 
conclusions remain the same: the coefficients $\gamma_n(\delta)$ are strongly affected by the choice of the
convention for $\delta$ and, whereas for none of the systems considered here the curvature-dependence of the interfacial 
tension is exactly in agreement with morphometric thermodynamics,
the morphometric form of the interfacial tension may be an
excellent approximation for suitable conventions for $\delta$.

In order to further assess to which extent
the above findings are generic, an additional
excess part $V$ of the external potential $V^\text{ext}$ is considered. 
This excess part $V$ is obtained by integrating the pair potential
$U_w(r)$ between a fluid particle and a wall particle over the volume $\tilde{\mathcal{V}}$
of the wall:
\begin{align}
  \beta V(\mathbf{r})=\varrho_w\int\limits_{\tilde{\mathcal{V}}}d^3r'\beta
  U_w(|\mathbf{r}-\mathbf{r'}|),
  \label{eq:soft_part}
\end{align}
where in $\tilde{\mathcal{V}}$ the number density $\varrho_w$ of the wall is taken to be constant.
The particle-wall potentials $U_w(r)$ are chosen to be of a similar form as the
pair potentials $U(r)$ between the fluid particles (Eqs.~(\ref{eq:pp-sw})-(\ref{eq:pp-lj})),
with the exception that the Yukawa-like wall-particle potential is not
truncated and the repulsive part of a 
Lennard-Jones-like wall potential is replaced by a hard wall. The pair
potentials are characterized by an energy scale $U_{0w}$, a length scale
$L_w$ and a cut-off length $L_{cw}$ such that $U_w(r>L_{cw})=0$:
\begin{itemize}
   \item the square-well ($U_{0w}<0$) or square-shoulder ($U_{0w}>0$) potential
         \begin{align}
            \beta U_w(r\leq L_{cw})=\beta U_{0w}
            \label{eq:pw-sw}
         \end{align}
         with $L_w=L_{cw}$,

   \item the Yukawa potential ($U_{0w}>0$, $L_{cw}=\infty$)
         \begin{align}
            \beta U_w(r)=\beta U_{0w}
                       \frac{L_w}{r}\exp\left(-\frac{r}{L_w}\right),
           \label{eq:pw-yu}
         \end{align}

   \item the Lennard-Jones potential ($U_{0w}>0$)
         \begin{align}
            \beta U_w(r)=
                   \begin{cases}
                       \beta U_{0w}
                       \left(\left(\frac{L_w}{r}\right)^{12}-\left(\frac{L_w}{r}\right)^6\right)&,L_w\leq
                       r\leq L_{cw}\\
                       \infty&,r<L_w.
                   \end{cases}
            \label{eq:pw-lj}
         \end{align}
\end{itemize}
The resulting excess parts $V$ of the substrate potentials as function of the radial distance $z$ to a
spherical reference surface of radius $R$ are given in Eqs.~(\ref{eq:soft_part_sw}),
(\ref{eq:soft_part_yu}), and (\ref{eq:soft_part_lj}) in Appendix \ref{sec:soft_parts}.

In Fig.~\ref{fig:intdspl_U0tilde_0.1_Vp_0.1} the results for systems with a
non-vanishing excess part $V$ of the external potential are shown. The
parameters are chosen such that, apart from $V\neq0$, the same systems as in
Figs.~\ref{fig:intdspl_mu_-3.95_U0tilde_0.1_Lc_1._pp_sw}(a),
\ref{fig:intdspl_mu_-3.94_U0tilde_0.1_Lc_5._pp_yu}(a), and
\ref{fig:intdspl_mu_-3.92_U0tilde_0.1_Lc_5._pp_lj}(a) are analyzed. It again
turns out that the coefficients $\gamma_n(\delta)$ exhibit a strong dependence on $\delta$ so
that the quality of morphometric thermodynamics as an approximation depends sensitively on
the convention for $\delta$. As a general trend, for all three
examples shown in Fig.~\ref{fig:intdspl_U0tilde_0.1_Vp_0.1} the non-morphometric
coefficient $|\beta\gamma_3(\delta)|L^2$ is not negligible in wider ranges of
conventions for $\delta$ than in the cases without an excess
substrate potential $V$. In this sense the quality of morphometric
thermodynamics as an approximation deteriorates in the presence of excess parts $V$
of the substrate potential.

\section{\label{section:conclusion_and_summary}Conclusions and Summary}
By using density functional theory within the second virial approximation we have
analyzed several model fluids with small number densities in contact with planar, spherical, or cylindrical
walls. We have compared the curvature expansion in Eq.~(\ref{eq:curvature_expansion})
of the interfacial tension $\gamma$
with the expression derived within morphometric thermodynamics
(Eq.~(\ref{eq:surface_tension_hadwiger_in_curvature})). Particular attention has been paid
to the implications of the choice of the position of the interface, which underlies the definition of the interfacial 
tension (Eq.~(\ref{eq:surface_tension_with_delta}) and Fig.~\ref{fig:delta}).
For none of the considered systems the expression for
the interfacial tension in accordance with morphometric thermodynamics
is exact, regardless of whether the
particles interact with each other via a square-well or square-shoulder potential
(Eq.~(\ref{eq:pp-sw})), a Yukawa potential (Eq.~(\ref{eq:pp-yu})), or a
Lennard-Jones potential (Eq.~(\ref{eq:pp-lj})).
As shown in
Figs.~\ref{fig:intdspl_mu_-3.88_U0tilde_1e300_Lc_1._pp_sw}-\ref{fig:intdspl_mu_-3.92_U0tilde_0.1_Lc_5._pp_lj}
the coefficients $\gamma_n(\delta)$ of the curvature expansion in
Eq.~(\ref{eq:curvature_expansion}) may depend sensitively on the chosen interface convention,
which is expressed in terms of the shift parameter $\delta$ (Fig.~\ref{fig:delta}). There are
conventions for which the morphometrically allowed coefficients are much larger than the
morphometrically forbidden ones so that within them morphometric thermodynamics is a reliable
approximation of the interfacial tension. However, the opposite situation can occur for
other interface conventions, in which case morphometric thermodynamics has to be used with
caution. In particular the reliability of morphometric thermodynamics as an
approximation deteriorates in the presence of excess contributions to
the wall potential (Fig.~\ref{fig:intdspl_U0tilde_0.1_Vp_0.1}).
Based on these results, it turns out to be necessary in future applications of morphometric thermodynamics to
clearly state which interface convention is chosen and why morphometric thermodynamics is expected to be a reliable
approximation for that particular interface convention as compared with others.


\appendix

\section{\label{sec:ideal_gas}Ideal Gas}
In this appendix we analyze the exactly solvable case of non-interacting particles. The density
functional in Eq.~(\ref{eq:density_functional}) with $U(r)=0$ is minimized 
by the equilibrium number density
\begin{align}
  \begin{aligned}
    \varrho_\text{eq}(\mathbf{r})&=\varrho^\text{bulk}_\text{eq}\exp\left(-\beta
      V^\text{ext}(\mathbf{r})\right)\quad\text{with}\\
      \varrho^\text{bulk}_\text{eq}&=\Lambda^{-3}\exp(\beta\tilde{\mu}).
  \end{aligned}
  \label{eq:density_ideal_gas}
\end{align}
For pointlike ideal gas particles the convention $\delta=0$ is convenient
and will be used throughout this appendix. For this choice the interface of area $\mathcal{A}$,
the reference surface, and the geometrical wall surface are the
same and the interfacial tension $\gamma$ is given by
\begin{align}
  \beta\gamma=-\frac{\varrho^\text{bulk}_\text{eq}}{\mathcal{A}}\int_\mathcal{V}\d^3r\left(\exp(-\beta
  V(\mathbf{r}))-1\right),\delta=0.
  \label{eq:interfacial_tension_ideal_gas}
\end{align}
The integration volume $\mathcal{V}$ in
Eq.~(\ref{eq:interfacial_tension_ideal_gas}) equals the volume accessible to
the fluid particles. Therefore the integrand depends only on the excess part $V$ of the external
potential $V^\text{ext}$ (see Eqs.~(\ref{eq:planar_external_potential}) and
(\ref{eq:spherical_external_potential})).\\
In the case of a hard wall with $V=0$ the interfacial tension of the ideal gas is zero, $\gamma=0$,
irrespective of the shape of the wall. Therefore the ideal gas is a useful
choice for studying the influence of the excess part $V\neq0$ of an external potential on the
morphometric coefficients.\\
Here we analyze a \textit{Yukawa-like} interaction $U_w$ (Eq.~(\ref{eq:pw-yu})) between
the fluid particles and the
wall particles.
The excess part $V$ follows from Eq.~(\ref{eq:soft_part}). For
planar, spherical, and cylindrical walls, respectively, one finds
\begin{align}
  \begin{aligned}
    \frac{V(z)}{V_p^{yu}}=&\exp\left(-\frac{z}{L_w}\right),\quad\text{plane},\\
    \frac{V(r)}{V_p^{yu}}=&\frac{L_w}{r}\exp\left(-\frac{r}{L_w}\right)\\
                     &\times\left[(S-1)\text{e}^S+(S+1)\text{e}^{-S}\right],\quad\text{sphere},\\
    \frac{V(r)}{V_p^{yu}}=&2SI_1(S)K_0\left(\frac{r}{L_w}\right),\quad\text{cylinder},
  \end{aligned}
  \label{eq:yukawa_soft_parts}
\end{align}
where $V_p^{yu}=2\pi\varrho_wU_{0w}L_w^3$ denotes the strength of the excess part $V$ at contact with a \textit{p}lanar wall and
$S=R/L_w$ with $K_0$ and $I_1$
as familiar modified
Bessel functions. The expressions in Eq.~(\ref{eq:yukawa_soft_parts}) for the
excess parts $V$ of the external potential facilitate to determine exactly the
coefficients $\gamma_n$ of the curvature expansion of the interfacial tension $\gamma$ in
Eq.~(\ref{eq:interfacial_tension_ideal_gas}). In the case of a spherical wall we obtain
\begin{align}
  \begin{aligned}
    \beta\gamma=-\varrho_\text{eq}^\text{bulk}L_w\sum\limits_{n=1}^\infty\frac{(-\beta
      V_p^{yu})^n}{n!\,n}\bigg\{1+\frac{L_w}{R}\left[\frac{2}{n}-1-n\right]\\
        +\frac{L_w^2}{R^2}\left[\frac{n^2}{2}+\frac{n}{2}-1-\frac{3}{n}+\frac{2}{n^2}\right]\\
        -\frac{L_w^3}{R^3}(2-n)(1-n)\left[\frac{1}{n^2}+\frac{1}{n}+\frac{1}{2}+\frac{n}{6}\right]\\
        +O\left(\frac{L_w^4}{R^4}\right)\bigg\}+O\left(\exp\left(-2\frac{R}{L_w}\right)\right),\delta=0,
  \end{aligned}
  \label{eq:surface_tension_id_gas_spherical}
\end{align}
and for the cylindrical wall the corresponding result is given by
\begin{align}
  \begin{aligned}
    \beta\gamma=-\varrho_\text{eq}^\text{bulk}L_w\sum\limits_{n=1}^\infty\frac{(-\beta
      V_p^{yu})^n}{n!\,n}\bigg\{1+\frac{L_w}{R}\left[\frac{1}{n}-\frac{1}{2}-\frac{n}{2}\right]\\
        +\frac{L_w^2}{R^2}\left[-\frac{1}{2n}-\frac{1}{8}+\frac{n}{8}+\frac{n^2}{8}\right]+O\left(\frac{L_w^3}{R^3}\right)\bigg\},\delta=0.
  \end{aligned}
  \label{eq:surface_tension_id_gas_cylindrical}
\end{align}
The expression for the planar wall is included in the expressions
for the curved walls (Eqs.~(\ref{eq:surface_tension_id_gas_spherical}) and
(\ref{eq:surface_tension_id_gas_cylindrical})) as the term being independent of the
radius $R$.\\
In Eqs.~(\ref{eq:surface_tension_id_gas_spherical}) and
(\ref{eq:surface_tension_id_gas_cylindrical}) the respective curvature expansions
are presented up to and including the leading \textit{non}-morphometric coefficients
(belonging to $R^{-3}$ in the case of spherical walls and to $R^{-2}$ in the case of
cylindrical walls) which in general are non-zero.\\
Further interesting insight can be gained by studying ratios of
particular coefficients:
\begin{itemize}
  \item The ratio of the leading coefficients $\gamma_1$ (belonging to $R^{-1}$),
    \begin{align}
      \frac{\gamma_1^s}{\gamma_1^c}=2,
    \end{align}
    i.e., a constant value independent of the
    strength $\beta V_p^{yu}$ of the external potential.
  \item For small amplitudes $\beta V_p^{yu}$ the ratio of the subdominant
    coefficients $\gamma_2$ (belonging to $R^{-2}$) is given by
    \begin{align}
      \frac{\gamma_2^s}{\gamma_2^c}=\frac{8}{3}\left(1-\frac{1}{18}\left(\beta
        V_p^{yu}\right)^2+O\left(\left(\beta V_p^{yu}\right)^3\right)\right).
      \label{eq:second_order_coefficients_ideal_gas}
    \end{align}
     For $\beta V_p^{yu}\ll1$ this ratio reduces to the constant value $8/3$ which has also been found in
     Ref.~\cite{Urrutia2014} and in the numerical calculations of Sec. \ref{section:discussion}.
  \item When comparing the leading ($\gamma_1^c$) and the subdominant ($\gamma_2^c$)
     coefficients for cylindrical walls we obtain
    \begin{align}
      \frac{\gamma_1^c}{\gamma_2^c}=-\frac{2}{3}\beta V_p^{yu}+\frac{67}{162}\left(\beta
      V_p^{yu}\right)^2+O\left(\left(\beta V_p^{yu}\right)^3\right).
    \end{align}
    This implies that for $\beta V_p^{yu}\ll1$ one has
    $|\gamma_2^c|\gg|\gamma_1^c|$ which
    contradicts the morphometric prediction according to which $\gamma_2^c$ should be zero.
\end{itemize}
These results for an ideal gas of non-interacting particles invalidate
morphometric thermodynamics if
there is a non-vanishing excess part $\beta V_p^{yu}\neq0$ of the external potential.

\begin{widetext}
\section{\label{sec:soft_parts}Excess parts of the external potentials}
The calculation of the excess parts $V$
of the external potentials
according to Eq.~(\ref{eq:soft_part})
results in lengthy expressions. Here these are presented for spherical walls
as function of the distance $z\geq0$ from the reference surface of radius
$R$, i.e., the distance from the center of the spherical wall is $r=R+z$.\\
In the case of the square-well or square-shoulder potential (Eq.~(\ref{eq:pw-sw}))
the excess part $V$ of the external potential is given by
  \begin{align}
   \begin{aligned}
    \beta V(z\geq0)&=\beta V_p^{sq}\times
    \begin{cases}
      z\geq L_w:&0\\
      z\leq L_w-2R:&2\frac{R^3}{L_w^3}\\
      L_w-2R<z<L_w:&\frac{1}{8}\frac{L_w^3}{(R+z)^3}\bigg[\frac{z^6}{L_w^6}+6\frac{z^5R}{L_w^6}+\frac{z^4}{L_w^4}\left(9\frac{R^2}{L_w^2}-6\right)\\
                  &+\frac{z^3}{L_w^3}\left(4\frac{R^3}{L_w^3}-24\frac{R}{L_w}+8\right)+\frac{z^2}{L_w^2}\left(-30\frac{R^2}{L_w^2}+24\frac{R}{L_w}-3\right)\\
                  &+\frac{z}{L_w}\left(-12\frac{R^3}{L_w^3}+24\frac{R^2}{L_w^2}-6\frac{R}{L_w}\right)+8\frac{R^3}{L_w^3}-3\frac{R^2}{L_w^2}\bigg],
    \end{cases}\\
    \beta V_p^{sq}&=\frac{2}{3}\pi L_w^3\beta U_{0w}\varrho_w.
   \end{aligned}
  \label{eq:soft_part_sw}
  \end{align}
In the case of the Yukawa potential (Eq.~(\ref{eq:pw-yu})) the excess part $V$ is given by
\begin{align}
  \begin{aligned}
    \beta V(z\geq0)=&\beta
    V_p^{yu}\exp\left(-\frac{z}{L_w}\right)\frac{L_w}{R+z}\left(\frac{R}{L_w}-1+\exp\left(-\frac{2R}{L_w}\right)\left(\frac{R}{L_w}+1\right)\right),\\
    \beta V_p^{yu}=&2\pi\varrho_w\beta U_{0w}L_w^3.
  \end{aligned}
  \label{eq:soft_part_yu}
\end{align}
(Note that Eq.~(\ref{eq:soft_part_yu}) and the sphere expression in Eq.~(\ref{eq:yukawa_soft_parts}) are identical.)\\
In the case of the Lennard-Jones-like potential (Eq.~(\ref{eq:pw-lj})) the reference
surface and the \textit{g}eometrical wall surface do not coincide: $R=R_g+L_w$ with
$R_g$ denoting the radius of the geometrical wall. Integration according to Eq.~(\ref{eq:soft_part}) results in an
expression $\beta V_g(z_g,R_g)$ in which $z_g$ measures the distance from the geometrical
wall surface, i.e., the
distance from the center of the spherical wall is $r=R_g+z_g$:
  \begin{align}
      \beta V_g(z_g,R_g)=&\beta
      V_p^{LJ}\frac{\mathcal{I}_{12}(z_g,R_g)-\mathcal{I}_{6}(z_g,R_g)}{-2\pi
        L_w^3\left\{-\frac{1}{10}\left(\frac{10}{9}\frac{L_w^9}{L_{cw}^9}-\frac{L_w^{10}}{L_{cw}^{10}}-\frac{1}{9}\right)+\frac{1}{4}\left(\frac{4}{3}\frac{L_w^3}{L_{cw}^3}-\frac{L_w^4}{L_{cw}^4}-\frac{1}{3}\right)\right\}},\notag\\
      \beta V_p^{LJ}=&-2\pi L_w^3\varrho_w\beta U_{0w}\left\{-\frac{1}{10}\left(\frac{10}{9}\frac{L_w^9}{L_{cw}^9}-\frac{L_w^{10}}{L_{cw}^{10}}-\frac{1}{9}\right)+\frac{1}{4}\left(\frac{4}{3}\frac{L_w^3}{L_{cw}^3}-\frac{L_w^4}{L_{cw}^4}-\frac{1}{3}\right)\right\},\notag\\
      \mathcal{I}_n(z_g,R_g)=&\frac{2\pi}{n-2}L_w^{n-1}\frac{L_w}{R_g+z_g}\sum_{j=1}^5A_j^{(n)}(z_g,R_g),\notag\\
      A_1^{(n)}(z_g,R_g)=&L_w^{4-n}\Bigg\{\frac{1}{n-3}\frac{R_g}{L_w}\left(\frac{z_g^{3-n}}{L_w^{3-n}}+\frac{(2R_g+z_g)^{3-n}}{L_w^{3-n}}\right)\notag\\
&+\frac{1}{(n-3)(n-4)}\left(-\frac{z_g^{4-n}}{L_w^{4-n}}+\frac{(2R_g+z_g)^{4-n}}{L_w^{4-n}}\right)\Bigg\},\notag\\
      A_2^{(n)}(z_g,R_g)=&L_w^{4-n}\Theta\left(R_g+z_g-L_{cw}\right)\Bigg\{\frac{1}{3-n}\frac{R_g}{L_w}\frac{(2R_g+z_g)^{3-n}}{L_w^{3-n}}-\frac{1}{2}\frac{L_{cw}^{2-n}}{L_w^{2-n}}\frac{R_g^2}{L_w^2}\notag\displaybreak\\
       &-\frac{1}{(4-n)(3-n)}\left(\frac{(2R_g+z_g)^{4-n}}{L_w^{4-n}}-\frac{(R_g+z_g)^{4-n}}{L_w^{4-n}}\right)\Bigg\},\notag\\
      A_3^{(n)}(z_g,R_g)=&L_w^{4-n}\Theta(2R_g+z_g-L_{cw})\Theta(L_{cw}-R_g-z_g)\Bigg\{\frac{1}{3-n}\frac{R_g}{L_w}\frac{(2R_g+z_g)^{3-n}}{L_w^{3-n}}\notag\\
       &-\frac{1}{2}\frac{L_{cw}^{2-n}}{L_w^{2-n}}\frac{R_g^2}{L_w^2}-\frac{1}{3-n}\frac{L_{cw}^{3-n}}{L_w^{3-n}}\frac{L_{cw}-R_g-z_g}{L_w}\notag\\
       &+\frac{1}{2}\frac{L_{cw}^{2-n}}{L_w^{2-n}}\frac{(L_{cw}-R_g-z_g)^2}{L_w^2}+\frac{L_{cw}^{4-n}-(2R_g+z_g)^{4-n}}{(3-n)(4-n)L_w^{4-n}}\Bigg\},\notag\\
      A_4^{(n)}(z_g,R_g)=&L_w^{4-n}\Theta(z_g-L_{cw})\Bigg\{\frac{1}{3-n}\frac{R_g}{L_w}\frac{z_g^{3-n}}{L_w^{3-n}}+\frac{1}{2}\frac{L_{cw}^{2-n}}{L_w^{2-n}}\frac{R_g^2}{L_w^2}+\frac{z_g^{4-n}-(R_g+z_g)^{4-n}}{(4-n)(3-n)L_w^{4-n}}\Bigg\},\notag\\
      A_5^{(n)}(z_g,R_g)=&L_w^{4-n}\Theta(R_g+z_g-L_{cw})\Theta(L_{cw}-z_g)\Bigg\{\frac{1}{3-n}\frac{L_{cw}^{3-n}}{L_w^{3-n}}\frac{R_g+z_g-L_{cw}}{L_w}\notag\\
&+\frac{1}{2}\frac{L_{cw}^{2-n}}{L_w^{2-n}}\frac{(R_g+z_g-L_{cw})^2}{L_w^2}
        +\frac{L_{cw}^{4-n}-(R_g+z_g)^{4-n}}{(4-n)(3-n)L_w^{4-n}}\Bigg\}.
    \label{eq:lj_vg}
  \end{align}
The excess part $V$ of the external potential as function of the distance $z$ from
the reference surface of radius $R$ is related to $V_g$ in Eq.~(\ref{eq:lj_vg}) via
\begin{align}
  \beta V(z\geq0)=\beta V_g(z+L_w,R-L_w).
  \label{eq:soft_part_lj}
\end{align}
\end{widetext}




\begin{thebibliography}{00}

   \bibitem{Koenig2004}
      P.-M.\ K\"{o}nig, R.\ Roth, and K.\ R.\ Mecke,
      Phys.\ Rev.\ Lett.\ \textbf{93}, 160601 (2004).

   \bibitem{Roth2006}
     R.\ Roth, Y.\ Harano, and M.\ Kinoshita,
     Phys.\ Rev.\ Lett.\ \textbf{97}, 078101 (2006).

   \bibitem{Goos2007}
     H.\ Hansen-Goos, R.\ Roth, K.\ Mecke, and S.\ Dietrich,
     Phys.\ Rev.\ Lett.\ \textbf{99}, 128101 (2007).

   \bibitem{Laird2012}
      B.\ B.\ Laird, A.\ Hunter, and R.\ L.\ Davidchack,
      Phys.\ Rev.\ E\ \textbf{86}, 060602(R) (2012).

   \bibitem{Evans2014}
     M.\ E.\ Evans and R.\ Roth,
     Phys.\ Rev.\ Lett.\ \textbf{112}, 038102 (2014).

   \bibitem{Hadwiger1957}
     H.\ Hadwiger,
     \textit{Vorlesungen \"{u}ber Inhalt, Oberfl\"{a}che und Isoperimetrie}
     (Springer, Berlin, 1957).

   \bibitem{Mecke1998}
     K.\ R.\ Mecke,
     Int.\ J.\ Mod.\ Phys.\ B\ \textbf{12}, 861 (1998).

   \bibitem{Bryk2003}
      P.\ Bryk, R.\ Roth, K.\ R.\ Mecke, and S.\ Dietrich,
      Phys.\ Rev.\ E\ \textbf{68}, 031602 (2003).

   \bibitem{Groh1999}
      B.\ Groh and S.\ Dietrich,
      Phys.\ Rev.\ E \textbf{59}, 4216 (1999).

   \bibitem{Blokhuis2013}
      E.\ M.\ Blokhuis,
      Phys.\ Rev.\ E\ \textbf{87}, 022401 (2013).

   \bibitem{Urrutia2014}
      I.\ Urrutia,
      Phys.\ Rev.\ E\ \textbf{89}, 032122 (2014).

   \bibitem{Goos2014}
     H.\ Hansen-Goos,
     J.\ Chem.\ Phys.\ \textbf{141}, 171101 (2014).

   \bibitem{Evans1979}
      R.\ Evans,
      Adv.\ Phys.\ \textbf{28}, 143 (1979).

   \bibitem{HansenMcDonald1976}
      J.\ P.\ Hansen and I.\ R.\ McDonald,
      \textit{Theory of Simple Liquids} (Academic, London, 1976).

   \bibitem{Schmidt2002}
      M.\ Schmidt, H.\ L\"{o}wen, J.\ M.\ Brader, and R.\ Evans,
      J.\ Phys.\ Condens.\ Matt. \textbf{14}, 9353 (2002).

   \bibitem{Rowlinson2002}
      J.\ S.\ Rowlinson and B.\ Widom,
      \textit{Molecular Theory of Capillarity} (Dover Publications, New York, 2002).

   \bibitem{Henderson2002}
      J.\ R.\ Henderson,
      J.\ Chem.\ Phys.\ \textbf{116}, 5039 (2002).


\end{thebibliography}
\end{document}